\documentclass{llncs}
\title{A  Tracer Driver for Versatile Dynamic Analyses of Constraint Logic Programs\thanks{In A. Serebrenik and S. Mu{\~n}oz-Hern{\'a}ndez (editors), Proceedings of the 15th Workshop on Logic-based methods in Programming Environments, October 2005, Spain. COmputer Research Repository (http://www.acm.org/corr/), cs.SE/0508105; whole proceedings: cs.PL/0508078.}\textsuperscript{,\,}\thanks{This work has been supported by the RNTL OADymPPaC project http://contraintes.inria.fr/OADymPPaC/}}

\author{Ludovic Langevine\thanks{This work has been partly carried out during the tenure of an ERCIM fellowship.}\inst{1} \and Mireille Ducassé\inst{2}} 

\institute{{\sc Sics} -- Uppsala Science Park, SE-75183 Uppsala, Sweden
\and {\sc Irisa/Insa} -- Campus universitaire de Beaulieu, F-35042 Rennes, France\\
\email{langevin@sics.se},\quad
\email{ducasse@irisa.fr}}

\usepackage{amsfonts}
\usepackage[T1]{fontenc}
\usepackage{times}
\usepackage{stmaryrd}
\usepackage{alltt}
\usepackage{tabularx}
\usepackage{multirow}
\usepackage{epsfig}
\usepackage{macro}

\usepackage[latin1]{inputenc}

\begin{document}
\maketitle
\begin{abstract}
Programs with constraints are hard to debug.  In this paper, we
describe a general architecture to help develop new debugging tools
for constraint programming.  The possible tools are fed by a single
general-purpose tracer.  A \emph{tracer-driver} is used to adapt the
actual content of the trace, according to the needs of the tool. This
enables the tools and the tracer to communicate in a client-server
scheme. Each tool describes its needs of execution data thanks to
\emph{event patterns}.  The tracer driver scrutinizes the execution
according to these event patterns and sends only the data that are
relevant to the connected tools.  Experimental measures show that this
approach leads to good performance in the context of constraint logic
programming, where a large variety of tools exists and the trace is
potentially huge.
\end{abstract}

\section{Introduction}

Program with constraints are especially hard to debug. The numerous
constraints and variables involved make the state of the execution
difficult to grasp. Moreover, the complexity of the filtering
algorithms as well as the optimized propagation strategies lead to a
tortuous execution. As a result, when a program gives incorrect
answers, misses expected solutions, or has disappointing performances,
the developer gets very little support from the current programming
environment to improve the program. This issue is critical because it
increases the expertise required to develop constraint programs.

Various work have addressed this critical issue. Most of them are
based on dynamic analyses. During the execution, some data are
collected in the execution so as to display some graphical views,
compute some statistics and other abstraction of the execution
behavior. Those data are then examined by the programmer to have a
better understanding of the execution. For instance, a display of the
search-tree helps to know how the search heuristics
behaves~\cite{fages02wlpe}. Adding some visual clues about the domain
propagation helps to see when the constraint propagation seems
inefficient~\cite{Outils:OPLstudio}. A more detailed view of the
propagation in specific nodes of the search-tree gives a good insight
to find out redundant constraints or select different filtering
algorithms. A common observation is that there is no ultimate tool,
that would meet all the debugging needs. There exists a large variety
of complementary tools, from coarse-grained abstraction of the whole
execution to very detailed views of small subparts, and even
application-specific displays.

The existing tools imply a dedicated instrumentation of the execution,
or a dedicated annotation of the traced program, to collect the data
they need. Those instrumentations are often hard to make and strongly
limit the use and the development of the tools. In this paper, we
present a generic approach where the possible tools are fed by a
single general-purpose tracer. A \emph{tracer-driver} is used to adapt
the actual content of the trace, according to the needs of the
tool. This enables the tools and the tracer to communicate in a
client-server scheme. Each tool describes its needs of execution data
thanks to
\emph{event patterns}.  The tracer driver scrutinizes the execution
according to these event patterns and sends only the data that are
relevant to the connected tools. A synchronisation mechanism allows
the tools to retrieve on demand more data about a given execution
event. Our experiments show that this architecture drastically reduces
the amount of trace data and significantly improves the performance.

Another description of the tracer driver focuses on the
architecture and implementation details, which are independent of the
traced programming language~\cite{langevine05aadebug}.
This paper focuses on the use of the tracer driver for CLP. Its main
contribution is an in-depth description of the good performance of the
approach, and especially of what is gained in the trace communication
and generation.

The paper is organized as follows.
Section~\ref{sec:analyse:tracer_driver} briefly presents the features of the
tracer driver. Section~\ref{sec:analyse:pattern} describes the event
patterns used to describe the needs of the
tools. Section~\ref{sec:analyse:request} lists the requests that an
analyzer can send to our tracer and how they are taken into account.
Section~\ref{sec:format} justifies the format used to
communicate the trace. Section~\ref{sec:analyse:performance} assesses
the performance of the scheme. Section~\ref{sec:related} discusses
related work.

\section[Pilote de traceur]{Overview of the tracer driver}
\label{sec:analyse:tracer_driver}
\index{traceur!pilote de}

This Section presents an overview of the tracer driver architecture
and, in particular, the interactions it enables between a tracer and 
analyzers. An analyzer is any tool that processes the trace.
The tracer and the analyzers are run at the same time.
%
Both synchronous and
asynchronous interactions are necessary between the tracer and the
analyzers. On the one hand, if analyzers need to get complements of
information at some events, it is important that the execution does
not proceed until the analyzers have decided so. On the other hand, if
the analyzers only want to collect information there is no need to
block the execution.

An execution trace is a sequence of observed execution events that
have attributes.  The analyzers specify the events to be observed by
means of \emph{event patterns}. An event pattern is a condition on
the attributes of an event (see details in
Sect.~\ref{sec:analyse:pattern}).  The tracer driver manages a base of
\emph{active} event patterns.  Each execution event is checked against
the set of active patterns.  An event matches an event pattern if and
only if the pattern condition is satisfied by the attributes of this
event.

An \emph{asynchronous pattern} specifies that, at matching trace events,
some trace data are to be sent to  analyzers without freezing the
execution.
A \emph{synchronous pattern} specifies that, at matching trace events,
some trace data are to be sent to analyzers. The execution is frozen
until the analyzers order the execution to resume. 
An \emph{event handler} is a procedure defined in an analyzer, that
is called when a matching event is encountered.

The architecture enables the management of several active patterns.
Each pattern is identified by a label.  A given execution event may
match several patterns. When sending the trace data to the analyzers
the list of (labels of) matched patterns is added to the trace. Then,
the analyzer mediator calls specific handlers for each matched
pattern and dispatches relevant trace data to them.  If at least one
matched pattern is synchronous, the analyzer mediator waits for every
synchronous handler to finish before sending the resuming command to
the tracer driver.  From the point of view of a given event handler,
the activation of other handlers on the same execution event is
transparent. Further details about this architecture can be found
in Langevine and Ducassé~\cite{langevine05aadebug}.

This article emphasizes more the tracer driver than the analyzer
mediator. On the one hand, the design and implementation of the tracer
driver is critical with respect to response time. Indeed it is called
at each event and executions of several millions of events (see
Sect.~\ref{sec:analyse:performance}) are very common. Every overhead,
even the tiniest, is therefore critical. On the other hand, the
implementation of the analyzer mediator is easier and much less
critical because it is called only on matching events.

\section{Event patterns}
\label{sec:analyse:pattern}

As already mentioned, an event pattern is a condition on the
attributes of events.  It consists of a first order formula combining
elementary conditions on the attributes.
%
%
%
This section summarizes the format of the trace events, specifies the
format of the event patterns and gives examples of patterns. 
%
%

\subsection{Trace events}
\label{sec:analyse:trace_event_format}

The actual format of the trace events has \emph{no influence} on the
tracer driver mechanisms. The important issue is that events have
attributes and that some attributes are specific to the type of
events.  
The trace format that we use is dedicated to constraint programming
over finite domains, formally defined in~\cite{langevine03}.

There are 15 possible 
event types in the tracer we use (choice-point, failure, solution, back-to,
new-variable, new-constraint, post, awake, reduce, suspend, entail, reject,
schedule, begin-exec, end-exec).  
%
Each event has common and specific attributes. The common attributes
are: the 
event type (called ``port''), a chronological event number, the depth
of the current node in the search-tree, the solver state (domains,
constraint store and propagation queue), and the \emph{user time}
spent since the beginning of the execution.  The specific attributes
depend on the port. For instance, a domain reduction event carries
data about the reduced variable (e.g. identifier and name), the
reducing constraint (e.g. external representation) and the removed
values.

\begin{figure}[t,t,t]
{\small
\begin{alltt}\tt
1 newVariable   v1=[0-268435455]
2 newVariable   v2=[0-268435455]
3 newConstraint c1 fd_element([v1,[2,5,7],v2])
4 reduce c1 v1=[1,2,3] W=[0,4-268435455]
5 reduce c1 v2=[2,5,7] W=[0-1,3-4,6,8-268435455]
6 suspend c1
\end{alltt}
\vspace{-0.5cm}
}\caption{A portion of trace for {\tt\small fd\_element(I,[2,5,7],A),
(A\#=I;A\#=2)}}
\label{fig:analyse:trace_example}
\end{figure}

Fig.~\ref{fig:analyse:trace_example} presents the beginning of a trace
of a toy program in order to illustrate the events described above.
This program 
%
specifies that {\tt A} is a finite domain variable which is in
$\{2,5,7\}$ and {\tt I} is the index of the value of {\tt A} in this
list; moreover {\tt A} is either equal to {\tt I} or equal to 2. The
second alternative is the only feasible one.
The trace can be read as follows. 
The first two events are related to the introduction of two variables
{\tt\small v}$1$ and {\tt\small v}$2$, corresponding respectively
to {\tt I} and {\tt A}.  In Gnu-Prolog, variables are always created
with the maximum domain (from 0 to 268.435.455).
Then the first constraint is created: {\tt fd\_element} (event \#3).
This constraint makes two domain reductions (events \#4 and \#5): the
domain of the first variable ({\tt I}) becomes $\{1,2,3\}$ and the
domain of {\tt A} becomes $\{2,5,7\}$. 
After these reductions, the constraint is suspended (event \#6).
%
The execution continues and finds the solution {\tt A=2,I=1}
through 32 other events not shown here.

\subsection[Motifs d'événements]{Patterns}
\label{sec:analyse:pattern_grammar}

\begin{figure}[t,t,t]
\begin{center}
\begin{small}
\begin{tabular}{p{1.5cm}p{0.5cm}l}
\nonterminal{pattern} &::= &\terminal{\em label}\terminal{:} \terminal{when}
\nonterminal{evt\_pattern} ~\nonterminal{op\_synchro}~
\nonterminal{action\_list}\\[0.05cm] 
\nonterminal{op\_synchro}&::=& \terminal{do}~ |~ \terminal{do\_synchro}\\[0.05cm]
\nonterminal{action\_list}&::=& \nonterminal{action}~ \terminal{,}~
\nonterminal{action\_list}~ |~ \nonterminal{action}\\[0.05cm]
\nonterminal{action}&::=&
\terminal{current(}\nonterminal{list\_of\_attributes}\terminal{)}~ 
|~ \terminal{call(}\nonterminal{procedure}\terminal{)}\\[0.3cm]
\nonterminal{evt\_pattern} &::= &\nonterminal{evt\_pattern}~ \terminal{or}~ \nonterminal{evt\_pattern}\hfill{\bf(1)}\quad\quad\\
&&| \nonterminal{evt\_pattern}~ \terminal{and}~ \nonterminal{evt\_pattern}\hfill{\bf(2)}\quad\quad\\
&&| \terminal{not}~ \nonterminal{evt\_pattern}\hfill{\bf(3)}\quad\quad\\
&&| \terminal{(} \nonterminal{evt\_pattern} \terminal{)}\hfill{\bf(4)}\quad\quad\\
&&| \nonterminal{condition}\hfill{\bf(5)}\quad\quad\\[0.05cm]
\nonterminal{condition}&::=& \terminal{{\em attribute}}~
\nonterminal{op2}~ \terminal{{\em value}}~ 
|~ \nonterminal{op1}\terminal{(}\terminal{{\em attribute}}\terminal{)}~~|~~
\terminal{true}\\[0.05cm] 
\nonterminal{op2}&::=& \terminal{<} | \terminal{>}
| \terminal{=} | \terminal{$\backslash$=} | \terminal{>=} | \terminal{=<}
| \terminal{in} | \terminal{notin}
| \terminal{contains} | \terminal{notcontains}\\[0.05cm]
\nonterminal{op1}&::=& \terminal{isNamed}\\[0.05cm]
\nonterminal{value}&::=& \emph{integer} | \emph{domain} | \emph{string}\\[0.05cm]
\nonterminal{attribute}&::=&\terminal{vident}\,|\,\terminal{vname}\,|\,\terminal{cident}\,|\,\terminal{cname}\,|\,\terminal{port}\,|\,\terminal{vdom}\,|\,\terminal{delta}\,|\,\terminal{chrono}\,\\&&|\,\terminal{depth}\,|\,\terminal{time}\,|\,\terminal{stage}\,|\,\terminal{node}
\end{tabular}

\end{small}
\end{center}
\caption{Grammar of event patterns}
\label{fig:analyse:grammar_patterns}
\end{figure}

We use patterns similar to the path rules of Bruegge and
Hibbard~\cite{bruegge83}.
Fig.~\ref{fig:analyse:grammar_patterns} presents the grammar of
patterns. A pattern contains four parts: a label, an event pattern, a
synchronization operator and a list of actions.
An event pattern is a composition of elementary conditions using logical
conjunction, disjunction and negation. 
A synchronization operator tells whether the pattern is asynchronous
(\terminal{do}) or synchronous (\terminal{do\_synchro}).
An action specifies either to ask the tracer driver to collect
attribute values
(\terminal{current(}\linebreak\nonterminal{list\_of\_attributes}\terminal{)}),
or to ask the analyzer to call a procedure
\terminal{call(}\nonterminal{procedure}\terminal{)}. Such a
procedure is written in a language that the analyzer can
execute. This language is independent of the tracer driver.
An elementary condition concerns an attribute of the current event.

There are several kinds of attributes. Each kind has a specific set of
operators to build elementary conditions.
For example, most of the common attributes are integer (chrono, depth,
node label).  Classical operators can be used with those attributes:
equality , disequality ($\neq$), inequalities ($<$,
$\leq$, $>$ and $\geq$).
The \emph{port} attribute has a set of 15 possible values. The
following operators can be used with the port attribute: equality and
disequality ($=$ and $\neq$) and two set operators, \terminal{in} and
\terminal{notin}.
Constraint solvers manipulate a lot of constraints and variables. Often,
a trace analysis is only interested in a small subset of them. Operators
\terminal{in} and \terminal{notin}, applied to identifiers of entities
or name of the variables, can specify such subsets. Operators
\terminal{contains} and \terminal{notcontains} are used to express
conditions on domains.

\subsection{Examples of patterns}
\label{sec:analyse:pattern_example}

\begin{figure}[t,t,t]
\begin{footnotesize}
\begin{alltt}
visu_tree:  when port in [choicePoint,solution,failure,backTo]
    do current(port=P and node=N and time=T),
       call buildTree(P,N,T)
new_cstr:   when port=newConstraint and stage='labeling'
    do current(cstrRep=Constraint),
       call recordDecision(Constraint)
visu_prop1: when port=reduce do current(vident=V and cident=C),
       call countReduce(V,C)
visu_prop2: when port=awake do current(cident=C),
       call countAwake(C)
synchronize: when port in [solution,failure]
    dosynchro refreshViewer(void)
\end{alltt}
\end{footnotesize}
\vspace{-0.5cm}
\caption{Examples of  patterns for visualization and monitoring}
\label{fig:analyse:patterns}
\end{figure}

\begin{figure}[t,t,t]
\begin{center}
\begin{minipage}[c]{0.55\linewidth}\includegraphics[width=\linewidth]{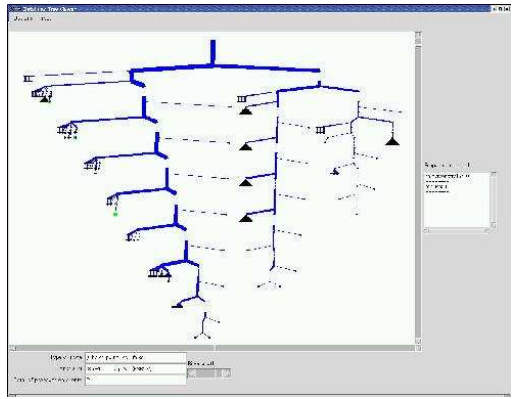}\end{minipage}
\caption{A display based on the patterns of Fig.~\ref{fig:analyse:patterns}.
}\label{fig:analyse:pavot}
\end{center}
\end{figure}

Fig.~\ref{fig:analyse:patterns} presents
five patterns that can be activated in parallel. These patterns aim at
producing a more or less precise view of the search-tree. Following
the user's parametrization, some of these patterns can be disabled, so
as to tune the trace volume. The first pattern ({\tt visu\_tree})
simply asks for a trace of the search-tree events: declaration of
choice-points, of leafs and of backtrackings. This is enough to
compute the structure of the search-tree. The second pattern ({\tt
new\_cstr}) adds to the trace the posting of every decision constraint
(a constraint that is posted by the labeling procedure).  It allows
the edges of the search-tree to be labeled with the decision
constraints they represent. Those two patterns gives the basic data
for the search-tree viewer: they are always enabled when the viewer is
running.

The following two patterns are added when more detailed data about the
nodes are needed.  ({\tt visu\_prop1}) asks for the trace of every
domain reduction, with the identifiers of the reduced variable and the
reducing constraint.  ({\tt visu\_prop2}) is interested in every
constraint awakening, with the identifier of the awakened
constraint. The various combinations of these two patterns allow the
computation of statistics about the number of constraint reduction,
the number of awakening or the proportion of useful awakenings
(awakenings followed by at least one domain reduction) in each node.
Such statistics can be used to add visual clues on the search
tree. For instance, the size of the nodes or the width of the edges
can depend on one of those indicators.

The last pattern synchronizes the display with the execution.  The
execution is often running much faster than its visualization.  This
issue can be adressed by such a synchronization mechanism.
Fig.~\ref{fig:analyse:pavot} shows a screenshot of the viewer using
those five patterns according to the user needs\footnote{This viewer
is part of Pavot, a tool developed at INRIA Rocquencourt. http://contraintes.inria.fr/~arnaud/pavot/}. In this
configuration, the width of the row depends on the total number of
propagation events (domain reductions and constraint awakenings)
occurring in the subtree.


\section[Analyseur de trace]{Analyzer mediator}
\label{sec:analyse:request} 

The analyzer mediator processes the trace: it specifies to the tracer
driver what events are needed and may execute specific actions for
each type of relevant events. The analyzer can supervise several
analyses at a time.  Each analysis has its own purpose and uses
specific pieces of trace data. The independence of the concurrent
analyses is ensured by the mediator that centralizes the communication
with the tracer driver and dispatches the trace data to the ongoing
analyses.



The requests that an analyzer can send to the tracer driver are of three
kinds.
Firstly, the analyzer can ask for additional data about the current
event. 
%
Secondly, the analyzer can modify the event patterns to be checked by
the tracer driver (the \emph{active} patterns). 
%
Thirdly, the analyzer can notify the end of a synchronous
session. 


Primitive {{\tt current}} specifies a list of event attributes to
retrieve in the current execution event.
%
%
%
The tracer retrieves the requested pieces of data and sends them
to the mediator.\quad
%
{{\tt reset}} deletes all the active event patterns and their
labels. 
%
%
\quad
Primitive {{\tt remove}} deletes the active patterns whose labels are
specified in the parameter.
%
\quad
Primitive {{\tt add}} inserts, in the active patterns, the event patterns
  specified in the parameter, following the grammar described in
  Figure~\ref{fig:analyse:grammar_patterns}.
%
\quad
Primitive {{\tt go}} notifies the tracer driver that a synchronous session
  is finished. The traced execution will be resumed.


\begin{figure}[t,t,t]
\begin{footnotesize}
\begin{alltt}
step :-
  {\bf{}reset},
  {\bf{}add}([step:when true 
     dosynchro call(tracer_toplevel)]),
  {\bf{}go}.
skip_reductions :-
  {\bf{}current}(cstr = CId and port = P),
  {\bf{}reset},
  (  P == awake
  -> {\bf{}add}([sr:when cstr = CId and port 
    in [suspend,reject,entail] dosynchro call(tracer_toplevel)]),
  ;  {\bf{}add}([step:when true dosynchro call(tracer_toplevel)])),
  {\bf{}go}.
\end{alltt}
\end{footnotesize}
\vspace{-0.5cm}
\caption{Implementation of two tracing commands}
\label{fig:analyse:standard_tracing}
\end{figure}

Fig.~\ref{fig:analyse:standard_tracing} illustrates the use of the
primitives to implement two tracing commands.
Command {\tt step} enables to go to the very next event. It simply
resets all patterns and adds one pattern which will match any event
(the associated condition is always true). This pattern calls, in a
synchronous way, the tracer toplevel. Therefore, the tracer will call
the toplevel at each event, and the toplevel will be synchronized with
the execution: the user will be able to investigate the current state
of the execution before resuming the execution.
Command {\tt skip\_reductions} enables to skip the details of variable
domain reductions when encountering the awakening of a constraint. It
first retrieves the current port, if it is \awake{} it asks to go to
the suspension of this constraint: the possible domain reductions are
skipped. There, the user will, for example, be able to check the value
of the domains after all the reductions. If the command is called on
an event of other type it simply acts as {\tt step}, so the tracer
will stop on the very next event.

\section{A Suitable Trace Format}
\label{sec:format}

An execution can generate several millions of execution events per
second. Large pieces of data can be attached to each event. The tracer
driver filters this trace according to the needs of the
analyzers. In this section, we study the properties of a good
format for execution traces to be sent from the tracer to the
analyzers. We consider several issues: the ability of the format to
represent partial (filtered) trace, the volume of the encoded trace,
and processing easiness.

\begin{figure}[t]
\begin{footnotesize}
\begin{alltt}
<reduce time="1045" vident="v13" />\\[0.15cm]<reduce cident="c12" vident="v13" />\\[0.15cm]<reduce chrono="1145697" time="1045" cident="c12" vident="v13"
        cexternal="greaterEq(v13,v19)">
  <delta vident="v13"><range from="0" to="21" /></delta>
  <update vident="v13" type="min" />
</reduce>
\end{alltt}
\end{footnotesize}
\caption{Three possible trace of the same event}
\label{fig:xmldomains}
\end{figure}

The tracer can access a large amount of data at each execution
event. Among those data, only a small subpart is in general needed for
a specific debugging tools. Therefore, the tracer driver only
communicates a small subset of the attributes and a part of the
current state. The concrete format has then to enable partial traces
to be represented, without losing the structure of the trace. An
instance of trace can thus be seen as an excerpt of the exhaustive
trace. The OADymPPaC project addressed this issue by designing an XML
format where most of the elements and attributes (in the sense of XML)
are optional~\cite{langevine05iclp}.  For instance,
Fig.~\ref{fig:xmldomains} presents three possible versions of the very
same execution event, a reduction of a domain.  The first version only
displays few basic attributes of the event: the user-time when the
event occurs and the identifier of the variable.  The second version
hides the user-time but displays also the identifier of the reducing
constraint, as well as some data about the values that have just been
removed. The third one is more complete: it displays the full set of
removed values, the external representation of the acting constraint
and the chronological number of the event. From an XML point-of-view,
they are three different excerpts of the same document. Of course,
the exhaustive document is never produced: the tracer driver only
fills the parts that are relevant according to the active patterns.

Since XML is a standard and
widely-used format, an interested developer can choose among dozens
of libraries to parse the trace data. Moreover, XML answers the
needs of trace structuring thanks to the notion of attributes and
nested elements. An event is an XML element that contains all its
attached data. It is worth noticing that there exists a standard
binary representation of XML: a table of symbols copes with the
verbosity of XML and speeds up the parsing of the trace~\cite{wbxml}.

\section{Experimental Results}
\label{sec:analyse:performance}

This section assesses the performances of the tracer driver and its
effects on the cost of the trace generation and communication.  It
shows several things. 
The overhead of the core tracer mechanisms is small. The
core tracer can thus be permanently activated.
The tracer driver overhead is acceptable and can be predicted.
There is no overhead for parallel search of patterns.
%
%
The tracer driver approach that we propose is more
efficient than sending over a default trace, even to construct
sophisticated graphical views.
Answering queries is orders of magnitude more efficient
than displaying traces.
There is no need to  a priori restrict the trace information.
The performance of our tool is comparable to the state-of-the-practice
while being more powerful and 
%
%
generic.


\subsection{Methodology of the Experiments}
\label{sec:performance:methodology}

When tracing a program, some time is spent in the program execution
($T_{prog}$), some time is spent in the core mechanisms of the
tracer\footnote{The core mechanisms include all the instructions that
are added to the traced execution such that the tracer can maintain
its own data. For instance, the generation of execution-unique
identifiers for variables, numbering the execution events or computing
the current depth in the search-tree are parts of those mechanisms.} 
($\Delta_{tracer}$), some time is spent in the tracer driver
($\Delta_{driver}$), some time is spent generating the requested trace
and sending it to the analysis process ($\Delta_{gcom}$), some time is
spent in the analyses ($\Delta_{ana}$). Hence, if we call $T$ the
execution time of a traced and analysed program, we approximatively
have: $T \simeq T_{prog} +
\Delta_{tracer} + \Delta_{driver} + \Delta_{gcom} + \Delta_{ana}$.


The mediator is a simple switch. The time taken by its execution is
negligible compared to the time taken by the simplest analysis,
namely the display of trace information. Trace analysis takes a time
which vary considerably according to the nature of the analysis. The
focus of this article is not to discuss which analyses can be achieved
in reasonable time but to show that a flexible analysis environment
can be offered at a low overhead. Therefore, in the following
measurements $\Delta_{ana}= 0$.

%
%

%
The experiments have been run on a PC, with a 2.4\,GHz
Pentium {\sc iv}, 512 Kb of cache, 1~GB of RAM, running under the
GNU/Linux~2.4.18 operating system.
%
The stable release (1.2.16) of \gprolog{} has been used.  The
tracer is an instrumentation of the source code of this very same
version and has been compiled in the same conditions by {\tt
  gcc-2.95.4}.
%
%
The execution times have been measured with the \gprolog{}
profiling 
predicates whose accuracy is 1\,ms. The measured executions consist of
a batch of executions such that each measured time is at least 20
seconds.
The measured time is the sum of \emph{system} and \emph{user} times.
%
%
Each experimental time given below is the average time of a series of
ten measurements.  In each series, the maximal relative deviation was
smaller than 1\,\%.

\subsection{Benchmark programs}
The 9 benchmark programs\footnote{Their source code is available at
http://contraintes.inria.fr/\~{}langevin/codeine/benchmarks} are
listed in Table~\ref{tab:analyse:characteristic_programs}, sorted by
increasing number of trace events.
  Magic(100), square(4), golomb(8) and golfer(5,4,4) are part of
  CSPLib, a benchmark library for constraints by
  Gent~and~Walsh~\cite{csplib}.
%
The golomb(8) program is executed with two strategies which exhibit
very different response times.
Those four programs have been chosen 
for their significant execution time and for the variety of
constraints they involve.
%
Four other programs have been added to cover more specific aspects of
the solver mechanisms:
Pascal Van Hentenryck's bridge problem (version of~\cite{gnuprolog});
%
two instances of the $n$-queens problem; and
%
``propag'', the proof of infeasibility of $1 \leq x, y \leq 70000000 \wedge x < y \wedge y < x$. 

\begin{table}[t]
\begin{center}
\begin{footnotesize}
\begin{tabular}{lrrrrrr}
{\bf Program}& evts ($10^6$)&Trace Size (Gb)&$T_{prog}$ (ns)&$\varepsilon$&$R_{tr.}$&Dev. for $T_x$\\
\hline
\emph{bridge}        & 0.2   & 0.1  & 14     & 72  & 1.21 & $\leq 0.4\%$\\
\emph{queens(256)}   & 0.8   & 1.5  & 173    & 210 & 1.14 & $\leq 0.2\%$\\
\emph{magic(100)}    & 3.2   & 1.4  & 215    & 66  & 1.03 & $\leq 0.2\%$\\
\emph{square(24)}    & 4.2   & 20.8 & 372    & 88  & 1.05 & $\leq 0.6\%$\\
\emph{golombF}       & 15.5  & 3.4  & 7,201  & 464 & 1.01 & $\leq 0.4\%$\\
\emph{golomb}        & 38.4  & 7.9  & 1,721  & 45  & 1.00 & $\leq 0.5\%$\\
\emph{golfer(5,4,4)} & 61.0  & >30  & 3,255  & 53  & 1.05 & $\leq 0.7\%$\\
\emph{propag}        & 280.0 & >30  & 3,813  & 14  & 1.28 & $\leq 1.0\%$\\
\emph{queens(14)}    & 394.5 & >30  & 17,060 & 43  & 1.08 & $\leq 0.4\%$\\
%
%

\end{tabular}
\end{footnotesize}
\end{center}
\caption{Benchmark Programs and tracer overhead}
\label{tab:analyse:characteristic_programs}
\end{table}

The benchmark programs have executions large enough for the
measurements to be meaningful. They range from 
200,000 events to about 400 millions events. 
Furthermore, they represent a wide range of CLP(FD) programs. 

The third column gives  the size of the traces of the benchmarked
programs for the default trace model.
All executions but the smallest one exhibit more than a gigabyte, for
executions sometimes less than a second. It is therefore not
conceivable to systematically generate such an amount of
information. As a matter of fact measuring these size took us 
hours and, in the last three cases, exhausted our patience!
Note that the size of the trace is not strictly proportional to the
number of events because the attributes collected at each type of
events are different. For example, for domain reductions, several
attributes about variables, constraints and domains are collected while
other types of events simply collect the name of the corresponding
contraint.

The fourth column gives $T_{prog}$, the execution time in $ms$ of the program
simply run by \gprolog{}.
The fifth column shows the average time of execution per event
$\varepsilon= \frac{T_{prog}}{\mathrm{Nb.\ evt.}}$. It is between 14 ns
and 464 ns per event. For most of the suite $\varepsilon$ is around
50ns. The three remarkable exceptions are \emph{propag} ($\varepsilon =
14$ ns), \emph{queens(256)} ($\varepsilon = 210$ ns) and \emph{golombF}
($\varepsilon = 464$ ns). The low $\varepsilon$ is due to the efficiency
of the propagation stage for the constraints involved in this
computation. The large $\varepsilon$s are due to a lower proportion of
``fine-grained'' events.





%
%

\paragraph*{Core tracer mechanisms can be permanently activated}
Table~\ref{tab:analyse:characteristic_programs} also gives the results
of the measurements of the overhead of the core tracer mechanisms.
Those results have already been discussed in~\cite{langevine03}.
For all the measured executions $R_{tracer}$ is less than 30\% in the
worst case, and less than 5\% for five traced programs.
%
%
That is are very positive. The
core mechanisms of the tracer can be systematically activated. Users
will hardly notice the overhead. Therefore, while developping
programs, users can directly work in ``traced'' mode, they do not need
to switch from untraced to traced environments. This is a great
confort.

\subsection{Tracer driver overhead}
%
The measure of 
$T_{driver}\simeq T_{prog} \!+\! \Delta_{core\_trace}\!+\!
\Delta_{driver}$ 
is the execution time of the program run by the
tracer  with the filtering procedure activated for generic
patterns. Only the attributes necessary for the requested patterns are
calculated at relevent events.  In order for $\Delta_{gcom}$ to be
zero, the patterns are designed such that no event matches them. One
run is done per pattern. The patterns are listed in
Figure~\ref{tab:analyse:patterns:driver}. Pattern {\bf 1a} is checked
on few events and on one costly attribute only. Pattern {\bf 2a} is
checked on numerous events and on two costly attributes. Pattern {\bf
3a} is checked on all events and on one cheap attribute. Pattern {\bf
4a} is checked on all events and systematically on three attributes.

\begin{figure}[t]
\footnotesize
 {\em {\bf 1a.} when port=post and isNamed(cname) 
do current(port,chrono,cident)}.

 {\em  {\bf 2a.} when port=reduce and 
(isNamed(vname) and isNamed(cname))\\
\quad\quad\quad do current(port,chrono,cident)}.

 {\em {\bf 3a.} when chrono=0 do current(chrono)}. 
  
  
 {\em {\bf 4a.} when depth=50000 or (chrono>=1 and node=9999999) 
do current(chrono,depth)}.
  
  {\em {\bf 5a}}: patterns 1a, 2a, 3a and 4a
    activated in parallel.
\caption{Patterns used to measure the tracer driver overhead}
\label{tab:analyse:patterns:driver}
\end{figure}

\begin{figure}[tttt]
\begin{center}
\includegraphics[width=0.67\linewidth]{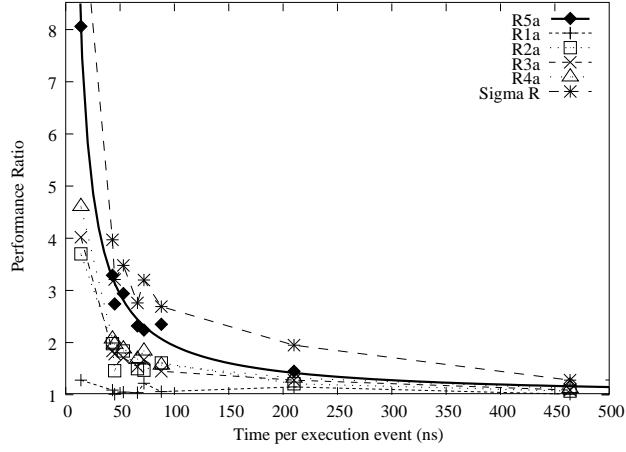}
\caption{Cost of the tracer driver mechanisms compared to
$\mathbf{\varepsilon}$ }
\label{tab:analyse:perf_bench}
\end{center}
\end{figure}


\paragraph*{Tracer driver overhead is acceptable}
Figure~\ref{tab:analyse:perf_bench} gives the results of the
measurements of the overhead of the tracer driver for all the benchmark
programs and for five patterns. The figure draws 
$R_{driver} = \frac{T_{driver}}{T_{prog}}$, 
compared to the average time per event ($\varepsilon$) for the 5
patterns. 
For all but one program, $R_{driver}$ is negligible for the very simple
patterns and less than 3.5 for pattern {\bf 5a} which is the
combination of 4 patterns. For programs with a large $\varepsilon$,
even searching for  pattern {\bf 5a} is negligible. In the worst case,
an overhead of 8 is still acceptable.
 
\paragraph*{No overhead for parallel search of patterns}
When $n$ patterns are checked in parallel they already save 
$(n-1)T_{tracer}$ compared to the search in sequence which requires
to executes $n$ times the program instead of one time.
Figure~\ref{tab:analyse:perf_bench} further shows that
\begin{center}
$\Delta_{driver}^{1a} + \Delta_{driver}^{2a} + \Delta_{driver}^{3a}
+ \Delta_{driver}^{4a} > \Delta_{driver}^{(1|2|3|4)a}$. 
\end{center}
As a matter of fact, the curve $\Sigma R = R_1+R_2+R_3+R_4-3$, that
adds the overheads of the four separated patterns, is above the curve
of $R^{5a}_{driver}$.  This means that not only is there no overhead
in the filtering mechanism induced by the parallel search, but there
is even a minor gain.

\paragraph*{Tracer driver overhead is predictable} 
The measured points of Figure~\ref{tab:analyse:perf_bench} can be
interpolated with curves of the form
$R_{driver} = a + b/\varepsilon$.  
%
%
This means that the tracer and tracer driver overheads per event can
be approximated to constants depending on the patterns and {\em
independant of the traced program.}

\subsection{Communication overhead}

The measure of 
$T_{gcom} \simeq T_{core\_tracer} + \Delta_{driver} + \Delta_{gcom}$ 
is the execution time of the program run by the tracer. A new set of
patterns are used so that some events match the patterns, the
requested attributes of the matched events are generated and sent to a
degenerated version of the mediator: a C-program that simply reads the
trace data on its standard input.
 Due to lack of space we only show the result
of program golomb(8) which has a median number of events and has a
median $\varepsilon$.

\begin{figure}[t]
\begin{small}
 {\em {\bf 1b.} cstr: when port=post
do current(chrono,cident,cinternal).} 
\\\phantom{{\em {\bf 1b.}}} 
{\em tree: when port in [failure,backTo,
choicePoint,solution] do current(chrono,node,port).} 

 {\em {\bf 2b.} newvar: when port=newVariable 
do current(chrono, vident, vname).}
\\\noindent\phantom{{\em {\bf 2b.}}} {\em dom: when port in [choicePoint,backTo,solution]}\\
\noindent\phantom{{\em {\bf 2b.}}}{\em do
current(chrono,node,port,named\_vars,full\_dom).}

 {\em {\bf 3b.} propag1: when port=reduce do current(chrono).}

 {\em {\bf 4b.} propag2: when port=awake do current(chrono).}
\end{small}
\caption[Motifs d'événements utilisés pour mesurer $
  \Delta_{com}$]{Event patterns used to assess the trace generation
  and the communication overhead}
\label{fig:analyse:patterns_com}
\end{figure}

The patterns are listed in
Figure~\ref{fig:analyse:patterns_com}.
Pattern {\bf 1b}, composed of two basic patterns, allows
a ``bare'' search tree to be constructed, as shown by most debugging tools. 
Pattern {\bf 2b} (two basic patterns) allows the display of 3D views of
variable updates as shown in Figure~\ref{fig:analyse:pavot}.
Pattern {\bf 3b} and
pattern {\bf 4b} provide two different execution details to decorate
search trees. Depending on the tool settings, three different visual
clues can be displayed. One is shown in
Figure~\ref{fig:analyse:pavot}.

\begin{table}[tttt]
\begin{center}\footnotesize
\begin{tabular}{|c||r|r|r|r|r|}
    \hline
  \multicolumn{6}{|c|}{Program: golomb(8)\qquad $\varepsilon=45\mathrm{ns}$\qquad $T_{prog=1.73\mathrm{s}}$}\\
  \hline
    Patterns & Traced evts $(10^6$)  & Trace size (Mb)    & Elapsed time (s) & $R_{dr.}$& $R_{gcom}$\\
  \hline
1b         & 0.36  &  21 & 4.50 & 1.03 & 2.6\\
2b         & 0.13  & 111& 16.17 & 1.02 & 9.35\\
3b         & 5.04  & 141& 33.57  & 1.14 & 19.40\\
4b         & 14.58 & 394& 89.40  & 1.32 & 51.68\\
\hline
(1|2)b     & 0.36  & 124 & 17.47  & 1.04 & 10.09\\
(1|3)b     & 5.40  & 162 & 36.08  & 1.15 & 20.85\\
(1|4)b     & 14.94 & 415 & 92.71  & 1.33 & 53.59\\
(1|3|4)b   & 19.97 & 556 & 122.72  & 1.44 & 70.93\\
(1|2|3|4)b & 19.97 & 660 & 136.80 & 1.44& 79.07\\
\hline
\emph{def. trace} 
           & 38.36 & 7,910  & 393.08& 1.96 & 227.21 \\
\hline


\end{tabular}
\caption{Cost of the trace generation and communication}
\label{tab:analyse:perf_bench_comm}
\end{center}
\end{table}

Table~\ref{tab:analyse:perf_bench_comm} gives the results for the
above patterns and some of their combinations. All combinations
correspond to existing tools. For example, combining {\bf 1b} with
{\bf 3b} or/and {\bf 4b} allows a Christmas tree as shown in
Figure~\ref{fig:analyse:pavot} to be constructed with two different
parameterizations.
%
The 2\textsuperscript{nd} column gives the
number of events which match the pattern.
The 3\textsuperscript{rd} column gives the size of the resulting XML trace
as it is sent to the tool.
The 4\textsuperscript{th} column gives the elapsed time\footnote{Here
  system and user time are not sufficient because two processes are at
  stake. $T_{prog}$ has been re-measured in the same conditions.}.
The 5\textsuperscript{th} column gives the ratio $R_{driver}$,
recomputed for each pattern.
The 6\textsuperscript{th} column gives the ratio $R_{gcom} =
\frac{T_{gcom}}{T_{prog}}$.

\paragraph*{Filtered trace is more efficient and more accurate than
  default trace}
The last line gives results for the \emph{default} trace.  The
\emph{default} trace contains twice as many events as the trace
generated by pattern {\bf (1|2|3|4)b}, but it contains more
attributes than requested by the pattern; As a result, its size is ten
times larger and its $R_{gcom}$ overhead is three times
larger.
%
%
As a consequence, the tracer driver approach that we propose is more
efficient than sending over a default trace, even to construct
sophisticated graphical views. The accuracy and the lower volume of the
trace ease its post-processing by debugging tools.

\paragraph*{Answering queries is more efficient
  than displaying traces} 
$R_{gcom}$ is always much larger that $R_{driver}$, from $2.6$ to
$79.07$ in our exemple. Therefore, queries using patterns that
drastically filter the trace have significantly better response time
than queries that first display the trace before analysing it.


\paragraph*{No need to  a priori restrict the trace information}
Many tracers limit a priori the trace information in order to reduce
the volume of trace. This restricts the range of possible dynamic
analyzes without preventing the big size and time overhead as shown
above with the default trace: it lacks some important information
while being huge.

With our approach, trace information which is not requested does
not cost much, therefore our trace model can afford to be much
richer. This enlarges the possibility of adding new dynamic analyses.

\paragraph*{Performance are comparable to the state-of-the-practice}
$R_{gcom}$ varies from 2.6 to 79.07. To give a comparison the
Mercury tracer of Somogyi and Henderson~\cite{somogyi99} is regularly
used by Mercury developers. For executions of size equivalent to
those of our measurements, the Mercury tracer overhead has been
measured from 2 to 15, with an average of 7~\cite{JD02tplp}.  Hence
the ratios for patterns {\bf 1b}, {\bf 2b} and {\bf 1|2b} are
quite similar to the state-of-the-practice debuggers. The other
patterns show an overhead that can discourage interactive
usage. However, these patterns are more thought of for monitoring
than debugging when the interaction does not have to be done in real
time. Note, furthermore, that for the measured programs, the absolute
response time is still on the range of two minutes for the worst
case. When debugging, this is still acceptable. 

Our approach allows therefore to have the tracer present but idle by
default. When a problem is encountered, simple queries can be set to
localize roughly the source of the problem. Then, more costly patterns
can be activated on smaller parts of the program. This is pretty much
like what experienced programmers do. The difference with our approach
is that they do not have to change tools, neither to reset the
parameterizations of the debugger.



\section{Related Work}
\label{sec:related}

Kraut~\cite{bruegge83} implements a finite state machine to find
sequences of execution events that satisfy some patterns (``path
rules''). Several patterns are allowed and they can be enabled or
disabled during the execution, using a labeling policy.  Specified
actions are triggered when a rule is satisfied but they are limited to
some debugger primitives, such as a message display or a counter
increasing.
%
%
The trace analysis is necessarily synchronous and cannot benefit from
the power of a complete programming language.

Reiss and Renieris~\cite{reiss01} have an approach similar to ours.
They also structure their dynamic analyses into three different
modules: 1) extraction of trace, 2) compaction and filtering and 3)
visualization. They provide a number of interesting compaction
functions which should be integrated in a further version of our
system. They, however, first dump the whole trace information in files
before any filtering is processed. With our tracer driver filtering is
done on the fly, and section~\ref{sec:analyse:performance} has shown
that this is much more efficient than first storing in
files.

Coca~\cite{ducasse99c} and Opium~\cite{ducasse99} provide a trace query
mechanism, respectively for C and Prolog. This mechanism is
synchronous and does not allow concurrent analyses. It can be easily
emulated with our tracer driver and an analyzer mediator written in
Prolog.

UFO~\cite{auguston02} offers a powerful language to specify
patterns and monitors. The patterns can involve several
events, not necessarily consecutive.  In our framework, the monitors
have to be implemented in the analyzer with a general programming
language. A further extension should allow at least to implement
monitors in the trace driver to improve efficiency.  UFO, however, does
not allow the same flexibility as our tracer driver, and is heavier to
use for interactive debugging.

%

A debugging library for Sicstus Prolog has been implemented by
Hanák et~al~\cite{hss04iclp}.  
No performance results are available.  Some
tuning of the trace display is possible but the tracer is based on
a complete storage of the trace and a postmortem investigation: this
is impractical with real-sized executions. The lazy
generation of the trace our tracer implements leads to the same kind
of trace data in a more efficient and practical way.

Some debugging tools enable the user to interact with the execution
states.  User acts on the current state of the execution to drive the
search-tree exploration (Oz Explorer~\cite{schulte97}), to add new
constraints on a partial solution (CLPGUI~\cite{fages02wlpe}), to
recompute a former state (both).  Those features are really helpful
but go much beyond the scope of this paper. Our approach is
complementary: it addresses the communication from the traced
execution to the debugging tools.





\section{Conclusion}

In this paper we presented a tracer driver which allows both
synchronous and asynchronous trace analysis in the same execution,
fitting all the needs of the classical usages of a tracer into a
single tool. We have defined an expressive language of event patterns
where relevant events are described by first order formul\ae{}
involving most of the data the tracer can access. Specific primitives
enable the retrieval of large pieces of data ``on demand'' and the
adaptation of the event patterns to the evolving needs of the trace
analyzer.  Therefore, the produced trace is accurate: trace
generation, trace communication and trace post-processing are speeded
up. As shown by the experiments, this approach leads to good
performance in the context of constraint logic programming, where a
large variety of tools exists and the trace is potentially huge.
The tracer driver provides a powerful front-end for complex debugging
tools based on trace data.

This approach has been prototyped in GNU-Prolog and is currently under
development within SICStus Prolog.
%
%
%
%






\begin{small}
\paragraph*{Acknowledgment}\quad
The authors thank Pierre Deransart and their OADymPPaC partners for
fruitful discussions, as well as  Guillaume Arnaud for
his careful beta-testing of Codeine.
\end{small}


\bibliographystyle{plain}
\bibliography{main}

\begin{thebibliography}{10}

\bibitem{auguston02}
M.~Auguston, C.~Jeffery, and S.~Underwood.
\newblock A framework for automatic debugging.
\newblock In W.~Emmerich and D.~Wile, editors, {\em Proceedings fo the 17th
  International Conference on Automated Software Engineering (ASE'02)}, pages
  217--222. IEEE Press, 2002.

\bibitem{Outils:OPLstudio}
C.~Bracchi, C.~Gefflot, and F.~Paulin.
\newblock Combining propagation information and search-tree visualization using
  {OPL} studio.
\newblock In A.~Kusalik, editor, {\em Proceedings of WLPE'01}, pages 27--39,
  Cyprus, D\'ecembre 2001. Computer Research Repository cs.PL/0111040.

\bibitem{bruegge83}
B.~Bruegge and P.~Hibbard.
\newblock Generalized path expressions: A high-level debugging mechanism.
\newblock {\em The Journal of Systems and Software}, 3:265--276, 1983.
\newblock Elsevier.

\bibitem{gnuprolog}
D.~Diaz.
\newblock {\sc Gnu} prolog, 2003.
\newblock http://gprolog.sourceforge.net/.

\bibitem{ducasse99c}
M.~Ducass\'{e}.
\newblock Coca: An automated debugger for {C}.
\newblock In {\em Proceedings of the 21st International Conference on Software
  Engineering}, pages 504--513. ACM Press, May 1999.

\bibitem{ducasse99}
M.~Ducass\'{e}.
\newblock Opium: An extendable trace analyser for {Prolog}.
\newblock {\em The Journal of Logic programming}, 39:177--223, 1999.
\newblock A. Bossi and Y. Deville (eds).

\bibitem{wbxml}
Ericsson, IBM, Motorola, and {Phone.com}.
\newblock {WAP} binary {XML} content format.
\newblock Note, W3C, http://www.w3.org/TR/wbxml/, June 1999.

\bibitem{fages02wlpe}
F.~Fages.
\newblock Clpgui: a generic graphical user interface for constraint logic
  programming over finite domains.
\newblock In A.~Tessier, editor, {\em Proc. of the 12th Workshop on Logic
  Programming Environments ({\sc Wlpe}'02)}. Computer Research Repository
  cs.SE/0207048, 2002.

\bibitem{csplib}
I.P. Gent and T.~Walsh.
\newblock {CSPLib}: a benchmark library for constraints.
\newblock Technical report, APES-09-1999, 1999.
\newblock Available from http://www.csplib.org/.

\bibitem{hss04iclp}
David Hanák, Tamás Szeredi, and Péter Szeredi.
\newblock Fdbg, the clp(fd) debugger library of sicstus prolog.
\newblock In B.~Demoen and V.~Lifschitz, editors, {\em Proc. of {\sc Iclp}'04},
  2004.
\newblock Poster. LNCS 3132.

\bibitem{JD02tplp}
E.~Jahier and M.~Ducass\'{e}.
\newblock Generic program monitoring by trace analysis.
\newblock {\em Theory and Practice of Logic Programming}, 2(4-5):611--643,
  July-September 2002.

\bibitem{langevine05aadebug}
L.~Langevine and M.~Ducassé.
\newblock A tracer driver for hybrid execution analyses.
\newblock In {\em AADEBUG'05}. ACM Press, 2005.

\bibitem{langevine03}
L.~Langevine, M.~Ducass\'e, and P.~Deransart.
\newblock A propagation tracer for {Gnu-Prolog}: from formal definition to
  efficient implementation.
\newblock In C.~Palamidessi, editor, {\em Proc. of the 19th Int. Conf. on Logic
  Programming}. Springer, LNCS 2916, 2003.

\bibitem{langevine05iclp}
Ludovic Langevine and the {OADymPPaC Team}.
\newblock Gentra4cp: a generic trace format for constraint programming.
\newblock In {\em 21th Int. Conf. on Logic Programming}, pages 433--434.
  Springer, 2005.
\newblock Poster.

\bibitem{reiss01}
S.P. Reiss and M.~Renieris.
\newblock Encoding program executions.
\newblock In M.-J. Harrold and W.~Schäfer, editors, {\em Proc. of the 23rd Int.
  Conf. on Software Engineering}. IEEE Press, 2001.

\bibitem{schulte97}
C.~Schulte.
\newblock Oz explorer: a visual constraint programming tool.
\newblock In L.~Naish, editor, {\em Proc. of the 14th Int. Conf. on Logic
  Programming}, pages 286--300. MIT Press, 1997.

\bibitem{somogyi99}
Z.~Somogyi and F.~Henderson.
\newblock The implementation technology of the {Mercury} debugger.
\newblock In {\em Proc. of the 10th WLPE}, volume 30(4). Elevier, ENTCS, 1999.

\end{thebibliography}

\end{document}